\documentclass[aps,prb,twocolumn,superscriptaddress,10pt]{revtex4-2}
\setcitestyle{numbers,square}
\pdfoutput=1

\usepackage[utf8]{inputenc}
\usepackage[T1]{fontenc}
\usepackage[english]{babel} 
\usepackage{amsmath, amsfonts, amssymb}
\usepackage[locale=US, separate-uncertainty=true, range-units=single, retain-unity-mantissa=false]{siunitx}
\usepackage{xspace}
\usepackage{nicefrac}
\usepackage{multirow}
\usepackage{physics}
\usepackage{graphicx}
\usepackage{bm}

\usepackage{enumerate}
\usepackage{hyperref}
\hypersetup{
	colorlinks = true,
	linkcolor = blue,
	citecolor = blue,
	filecolor = blue,
	urlcolor = blue
}

\usepackage{prettyref}
\newcommand{\pref}[1]{\prettyref{#1}}
\newrefformat{fig}{Fig.~\ref{#1}}
\newrefformat{tab}{Table~\ref{#1}}
\newrefformat{sec}{Sec.~\ref{#1}}
\newrefformat{app}{App.~\ref{#1}}
\newrefformat{eq}{Eq.~(\ref{#1})}
\DeclareSIUnit\angstrom{\text {Å}}
\DeclareSIUnit\fu{\text{f.u.}}

\newcommand{\KCF}{KCuF$_3$\xspace}
\newcommand{\KZF}{KZnF$_3$\xspace}
\newcommand{\KCZF}{KCu$_{1-x}$Zn$_x$F$_3$\xspace}
\newcommand{\CuFoct}{CuF$_6$\xspace}
\newcommand{\eg}{\ensuremath{{e_g}}\xspace}
\newcommand{\ttg}{\ensuremath{{t_{2g}}}\xspace}
\newcommand{\Pmtm}{\ensuremath{{Pm\bar3m}}\xspace}
\newcommand{\Ifmcm}{\ensuremath{{I4/mcm}}\xspace}
\newcommand{\xxyy}{\ensuremath{{x^2-y^2}}\xspace}
\newcommand{\zz}{\ensuremath{{z^2}}\xspace}
\newcommand{\diff}{\ensuremath{\mathrm{d}}\xspace}

\begin{document}

\title{Exploring energy landscapes of charge multipoles using constrained density functional theory}

\author{Luca Schaufelberger}
\affiliation{Materials Theory, ETH Z\"u{}rich, Wolfgang-Pauli-Strasse 27, 8093 Z\"u{}rich, Switzerland}
\author{Maximilian E. Merkel}
\email{maximilian.merkel@mat.ethz.ch}
\affiliation{Materials Theory, ETH Z\"u{}rich, Wolfgang-Pauli-Strasse 27, 8093 Z\"u{}rich, Switzerland}
\author{Aria Mansouri Tehrani}
\affiliation{Materials Theory, ETH Z\"u{}rich, Wolfgang-Pauli-Strasse 27, 8093 Z\"u{}rich, Switzerland}
\author{Nicola A. Spaldin}
\affiliation{Materials Theory, ETH Z\"u{}rich, Wolfgang-Pauli-Strasse 27, 8093 Z\"u{}rich, Switzerland}
\author{Claude Ederer}
\email{claude.ederer@mat.ethz.ch}
\affiliation{Materials Theory, ETH Z\"u{}rich, Wolfgang-Pauli-Strasse 27, 8093 Z\"u{}rich, Switzerland}

\date{\today}

\begin{abstract}
We present a method to constrain local charge multipoles within density-functional theory. Such multipoles quantify the anisotropy of the local charge distribution around atomic sites and can indicate potential hidden orders. Our method allows selective control of specific multipoles, facilitating a quantitative exploration of the energetic landscape outside of local minima. Thus, it enables a clear distinction between electronically and structurally driven instabilities.
We demonstrate the effectiveness of this method by applying it to charge quadrupoles in the prototypical orbitally ordered material KCuF$_3$. We quantify intersite multipole-multipole interactions as well as the energy-lowering related to the formation of an isolated local quadrupole. We also map out the energy as a function of the size of the local quadrupole moment around its local minimum, enabling quantification of multipole fluctuations around their equilibrium value. Finally, we study charge quadrupoles in the solid solution KCu$_{1-x}$Zn$_x$F$_3$ to characterize the behavior across the tetragonal-to-cubic transition.
Our method provides a powerful tool for studying symmetry breaking in materials with coupled electronic and structural instabilities and potentially hidden orders. 
\end{abstract}

\maketitle

\section{Introduction}

\begin{figure}
    \centering
    \includegraphics[width=1\linewidth]{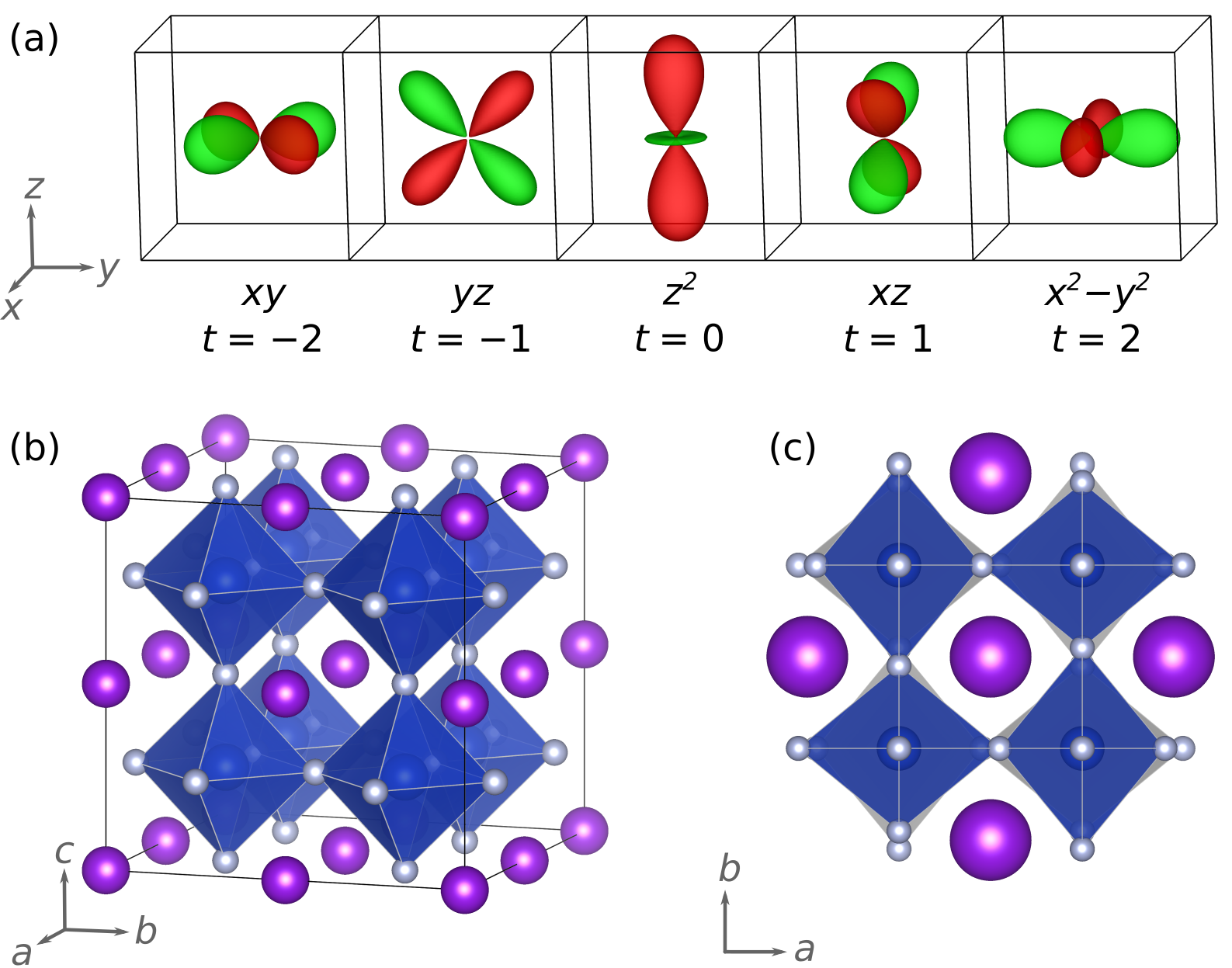}
    \caption{ (a) Visualization of the angular charge distribution corresponding to the five charge quadrupoles $Q_t \equiv w_{2t}$.
    Green and red regions represent regions with excess and reduced electronic charge relative to a spherical distribution, respectively.
    (b, c) Crystal structure of \KCF with K in purple, Cu in blue, and F in gray as well as blue \CuFoct octahedra.
    (b) High-symmetry cubic perovskite reference structure with \Pmtm symmetry.
    (c) Top view of the experimental crystal structure 
    \cite{tsukuda_stacking_1972} with \Ifmcm symmetry. The sign of the Jahn-Teller distortion alternates between adjacent \CuFoct octahedra, and is therefore described as a G-type antiferro distortion.
    }
\label{fig:crystal_struct_quads}
\end{figure}

Spontaneous symmetry breaking is ubiquitous in physics, found in fields ranging from cosmology and nuclear physics to condensed matter \cite{beekman_introduction_2019}. Prominent cases of spontaneous symmetry breaking in condensed matter physics are the emergences of charge or magnetic order \cite{coey_charge-ordering_2004, tchernyshyov_structural_2004, attfield_charge_2006, bordacs_magnetic-order-induced_2009}, which correspond to ordered arrangements of local charge monopoles or magnetic dipoles, respectively. Recently, more exotic forms of order, involving higher-order charge or magnetic multipoles have attracted considerable attention \cite{bultmark_multipole_2009, santini_multipolar_2009, spaldin_monopole-based_2013, suzuki_first-principles_2018, bhowal_revealing_2021, urru_neutron_2022}.
Higher-order multipoles encode anisotropies in the charge or magnetization density.
For example, in inversion symmetric materials, such anisotropy can be caused to leading order by charge quadrupoles, i.e., the second-order multipoles, depicted in \pref{fig:crystal_struct_quads}(a) as excess and depletion of electronic charges. Charge quadrupoles provide a physically intuitive framework for quantifying and analyzing the orbital order from spontaneous symmetry breaking \cite{mansouri_tehrani_untangling_2021, mansouri_tehrani_charge_2023}.

Higher-order multipoles are challenging to detect experimentally and thus constitute a type of ``hidden'' order \cite{aeppli_hidden_2020, pourovskii_hidden_2021}. The difficulty in experimental detection makes computational first-principles-based studies highly desirable.
In particular, in many materials, two or more order parameters are simultaneously involved in the symmetry breaking. For example, an electronic transition can occur simultaneously with a structural phase transition, such as orbital ordering together with a Jahn-Teller distortion \cite{tanaka_electron-density_1979, kugel_jahn-teller_1982, hirai_detection_2020, svoboda_orbital_2021, mansouri_tehrani_charge_2023}.
In such cases, it is not immediately clear which quantity is the primary order parameter driving the transition, and this question can be directly addressed in computational studies.

While varying the structural order parameters in simulations is commonly and easily done by constraining lattice parameters and atomic positions, varying the corresponding electronic order parameter for a fixed structure is challenging. This, however, is crucial for identifying the physical mechanisms driving spontaneous symmetry breaking and separating electronic from structural effects.
When exploring different local energy minima as a function of, e.g., arrangements of local charges or magnetic dipole moments, suitable initializations can be effective \cite{allen_occupation_2014, skokowski_influence_2020, malyi_local_2022, fiore_mosca_modeling_2022}. However, this becomes more cumbersome for higher-order multipole moments.
Additionally, it is often desirable to explore the physics beyond local minima and systematically vary the electronic order parameters both in the magnitude and the spatial arrangements of the local multipole moments.

Different approaches have been used to indirectly disentangle electronic and structural order parameters. For instance, by fitting a model to the results of first-principles calculations, the metal-insulator transitions in a range of rare-earth nickelates have been shown to be driven by proximity to an electronic instability which couples strongly to a lattice distortion~\cite{peil_mechanism_2019, georgescu_disentangling_2019, georgescu_quantifying_2022}.
Another approach has been pursued in studies on a variety of transition-metal perovskites with quadrupolar order, such as cuprates, manganates, vanadates, and titanates. Here, simulations separately varying the electronic temperature and frozen distortions show an electronic instability that does not require structural distortions but is often stabilized by them \cite{pavarini_mechanism_2008, pavarini_origin_2010, zhang_lavo3_2022}.

Here, we introduce a method for directly controlling the magnitude and spatial order of the electronic order parameter by constraining local multipoles independently of the crystal structure.
Our method applies the framework of constrained density-functional theory (DFT) \cite{dederichs_ground_1984, pickett_reformulation_1998} to local multipoles \cite{bultmark_multipole_2009} and enables the selective mapping of the energetic landscape as a function of local multipoles outside of local minima.
We introduce the methodology here for charge multipoles noting that it can easily be extended to magnetic and magnetoelectric multipoles \cite{bultmark_multipole_2009, thole_magnetoelectric_2018, bhowal_magnetoelectric_2022}.

While the main goal of the present manuscript is methodological, as a case study, we apply our method to \KCF, 
a prototypical Jahn-Teller-active perovskite with a $3d^9$ electronic configuration \cite{okazaki_polytype_1969, hutchings_neutron-diffraction_1969, tsukuda_stacking_1972}. It adopts a tetragonal structure with \Ifmcm symmetry and room-temperature lattice parameters of $a = b = \SI{4.14}{\angstrom}$ and $c = \SI{3.93}{\angstrom}$  \cite{knox_perovskite-like_1961, okazaki_polytype_1969, hutchings_neutron-diffraction_1969, marshall_unusual_2013}, which is stable up to at least \SI{800}{K} \cite{zhou_jahnteller_2011} and possibly even up to the decomposition temperature \cite{marshall_unusual_2013}.
This structure can be obtained from the high-symmetry cubic perovskite parent structure (\pref{fig:crystal_struct_quads}(b)) by applying an $R_3^+$ antiphase Jahn-Teller distortion of the \CuFoct octahedra together with a tetragonal distortion of the unit cell. 
The result is a G-type 3D-checkerboard antiferro-distortive pattern of alternating short and long Cu-F distances (\pref{fig:crystal_struct_quads}(c)).
At around \SI{38}{K}, \KCF becomes an A-type antiferromagnet, in which magnetic dipoles on the Cu ions order ferroically in the $a$-$b$-plane and antiferroically along the $c$ direction~\cite{hutchings_neutron-diffraction_1969}.

Numerous computational studies have investigated the orbital order and Jahn-Teller distortion in \KCF \cite{kugel_jahn-teller_1982, liechtenstein_density-functional_1995, pavarini_mechanism_2008, sims_thermally_2017, varignon_origins_2019}. For example, DFT+$U$ calculations have shown that even in its (hypothetical) high-symmetry cubic structure, \KCF has a tendency toward orbital ordering \cite{liechtenstein_density-functional_1995}, or equivalently in the language of multipoles, toward quadrupolar ordering \cite{svoboda_orbital_2021}.
This could suggest a purely electronic Kugel-Khomskii-type origin of the ordering due to superexchange~\cite{kugel_jahn-teller_1982, liechtenstein_density-functional_1995}.
However, DFT plus dynamical mean-field theory (DMFT) simulations showed that the Kugel-Khomskii superexchange mechanism leads to a hypothetical transition temperature of only around \SI{350}{K} in \KCF and structural distortions are necessary to stabilize the orbital order up to at least \SI{800}{K} \cite{pavarini_mechanism_2008}.
Thus, the symmetry breaking in \KCF appears to result from the interplay of electronic and structural degrees of freedom.

The remainder of this article is organized as follows:
First, we outline the derivation of multipole-constrained DFT.
We then verify our approach by applying it to \KCF and demonstrate its capabilities in the context of intersite quadrupole-quadrupole interactions. Finally, we use our method to study the electronic instability in the solid solution \KCZF across the transition from the tetragonal, Jahn-Teller-distorted \KCF to cubic \KZF, in which the Zn has a $3d^{10}$ configuration.

\section{Methods}

In this section, we first review the definition of charge multipoles \cite{bultmark_multipole_2009} and then sketch the derivation of the modification to the Kohn-Sham potential required to constrain them. Finally, we specify the parameters of the  DFT calculations used in this study.

\subsection{Charge multipoles}

To define charge multipoles, we consider the electron density $n(\bm r)$ in the local environment of an atom at the origin, $\bm r = \bm 0$, in an inhomogeneous external electric potential $\Phi(\bm r)$
\begin{align}
    \Phi(\bm r) =& \Phi(\bm 0) + r_i \partial_i \Phi(\bm r)|_{\bm r=0} + r_i r_j \partial_i \partial_j \Phi(\bm r)|_{\bm r=0} + \ldots,
    \nonumber
\end{align}
which has an electrostatic energy $E$ of
\begin{align}
    E / e =& -\int \diff \bm r n(\bm r) \Phi(\bm r) \nonumber\\
    =& - \left( \int \diff \bm r n(\bm r) \right) \Phi(\bm 0)
    - \left( \int \diff \bm r n(\bm r) r_i \right) \partial_i \Phi(\bm r)|_{\bm r=0} \nonumber\\
    &- \left( \int \diff \bm r n(\bm r) r_i r_j \right) \partial_i \partial_j \Phi(\bm r)|_{\bm r=0}
    - \ldots
    \label{eq:energy_expansion}
\end{align}
Here, the integrals in the parentheses correspond to charge multipoles of order $k=0$, 1, and 2, i.e., monopole, dipole, and quadrupole, respectively.

We now represent that electron density in terms of a density matrix $\rho_{\alpha'\alpha}$ in the basis of a complete set of atomic-like orbitals $\ket{\Psi_\alpha}$, such that
\begin{align}
    n(\bm r) = \expval{n}{\bm r} &= \sum_{\alpha\alpha'} \braket{\bm r}{\Psi_{\alpha'}}\mel{\Psi_{\alpha'}}{n}{\Psi_\alpha} \braket{\Psi_\alpha}{\bm r} \nonumber\\
    &= \sum_{\alpha\alpha'} \Psi_{\alpha'}(\bm r) \rho_{\alpha'\alpha} \Psi^*_\alpha(\bm r) \quad,
\end{align}
and use the fact that the atomic-like orbitals can be expressed as product of a radial part and a spherical harmonic, $\Psi_\alpha(r, \theta, \phi) = R_{nl}(r) Y_{lm}(\theta, \phi)$, where $n$ is the principal quantum number and $l$ and $m$ are the angular-momentum quantum numbers. 
Furthermore, the products of spatial coordinates in the integrals of a multipole of order $k$ in \pref{eq:energy_expansion} can be expressed in spherical harmonics, $Y_{kt}$, where $t \in \{-k, -k+1, \ldots, k\}$ labels the different types of multipoles of the same order $k$.
The integrals of \pref{eq:energy_expansion} thus separate into a radial and a (dimensionless) angular part. In the following, we only consider the angular part of the multipoles, in accordance with previous work~\cite{bultmark_multipole_2009, santini_multipolar_2009}.

For simplicity, we only consider terms of the density matrix that are diagonal in the angular momentum ($l=l'$), which can be assumed to give the dominant contributions for most relevant cases. We note that the generalization to $l \neq l'$ is straightforward. 
We can then express the $l$-contribution to the multipoles of order $k$ as
\begin{align}
    \hat w_{lkt} &\propto \sum_{mm'} \rho_{lm'lm} \int \diff\theta \diff\phi Y^*_{lm} Y_{lm'} Y_{kt} \sin\theta \nonumber\\
    &\propto \sum_{mm'} \rho_{lm'lm} \underbrace{(-1)^{m} \begin{pmatrix}
        l  & k & l \\
        -m & t & m'
    \end{pmatrix}}_{\propto \hat\mu^{lkt}_{mm'}} \quad, 
    \label{eq:pre-multipole}
\end{align}
which defines the \emph{charge-multipole operators} $\hat\mu^{lkt}$ up to some normalization. Here, the ``$2\times3$ matrix'' in \pref{eq:pre-multipole} is the Wigner 3-$j$ symbol, and, for ease of presentation, we have dropped all prefactors that depend only on $k$ and $l$.

In the following, we suppress the index $l$ (we will always focus on a single $l$ component) and instead add an index $I$ corresponding to the atomic site around which the multipoles are defined.
Furthermore, we use the normalization defined in Eq.~(26) of Ref.~\onlinecite{bultmark_multipole_2009},
which leads to the definition of charge multipoles we use in this manuscript:
\begin{align}
    w^I_{kt} &= \sum_{m,m'=-l}^l \mu^{kt}_{mm'} \rho^I_{m'm} \quad , \label{eq:dens_to_mult} \\
    \intertext{where $\mu$ is a real-valued linear combination of the complex-valued charge multipole operators:}
    \hat \mu^{kt}_{mm'} &= (-1)^{l-m+k}
    \begin{pmatrix}
        l & k & l \\
        -m & t & m'
    \end{pmatrix} n_{lk}^{-1} \nonumber \\
    \text{with}\quad
    n_{lk} &= (2l)! \left/ \sqrt{(2l-k)! (2l+k+1)!} \right. \quad .
    \label{eq:definition_mu}
\end{align}
Here, to obtain real-valued multipoles $w^I_{kt}$, we have used \pref{eq:sph_to_real} from \pref{app:multipoles_definition} to transform the multipole operators $\hat\mu^{kt}$ expressed in complex spherical harmonics to the $\mu^{kt}$ that are used in \pref{eq:dens_to_mult} and are expressed in real spherical harmonics. 
\pref{app:multipoles_definition} also contains further details on our definition of multipoles and the relation to the notation used in Ref.~\onlinecite{bultmark_multipole_2009}.
We note that $w^I_{kt}$, $\rho^I$, and $\mu^{kt}$ are dimensionless quantities since they contain only the angular part, while the full charge multipoles introduced in \pref{eq:energy_expansion} have units that depend on the specific multipolar order.

The multipole moments $w^I_{kt}$ defined in \pref{eq:dens_to_mult} can be easily obtained within DFT via the local density matrix (see, e.g., Refs.~\onlinecite{pickett_reformulation_1998, bengone_implementation_2000}),
\begin{align}
    \rho^I_{mm'} = \sum_{\bm p\nu} f_{\bm p\nu} \expval{P^I_{mm'}}{\psi_{\bm p\nu}} \quad.
    \label{eq:dens_in_dft}
\end{align}
Here, $|\psi_{\bm p \nu}\rangle$ and $f_{\bm p \nu}$ are the Kohn-Sham eigenfunctions and corresponding occupations at a wave vector $\bm p$ and for a band and spin indexed by $\nu$. The operators $P^I_{mm'} = | \Phi^I_{nlm'} \rangle \langle \Phi^I_{nlm}|$ correspond to the projection on a suitable local atomic-like basis at site $I$ ($n$ and $l$ indices suppressed for ease of notation), which is readily available in most DFT codes. This decomposition of a charge density into its multipole components is schematically illustrated in the right part of \pref{fig:schematic_method}.

\subsection{Constraining multipoles in DFT}

\begin{figure*}
    \centering
    \includegraphics[width=1\linewidth]{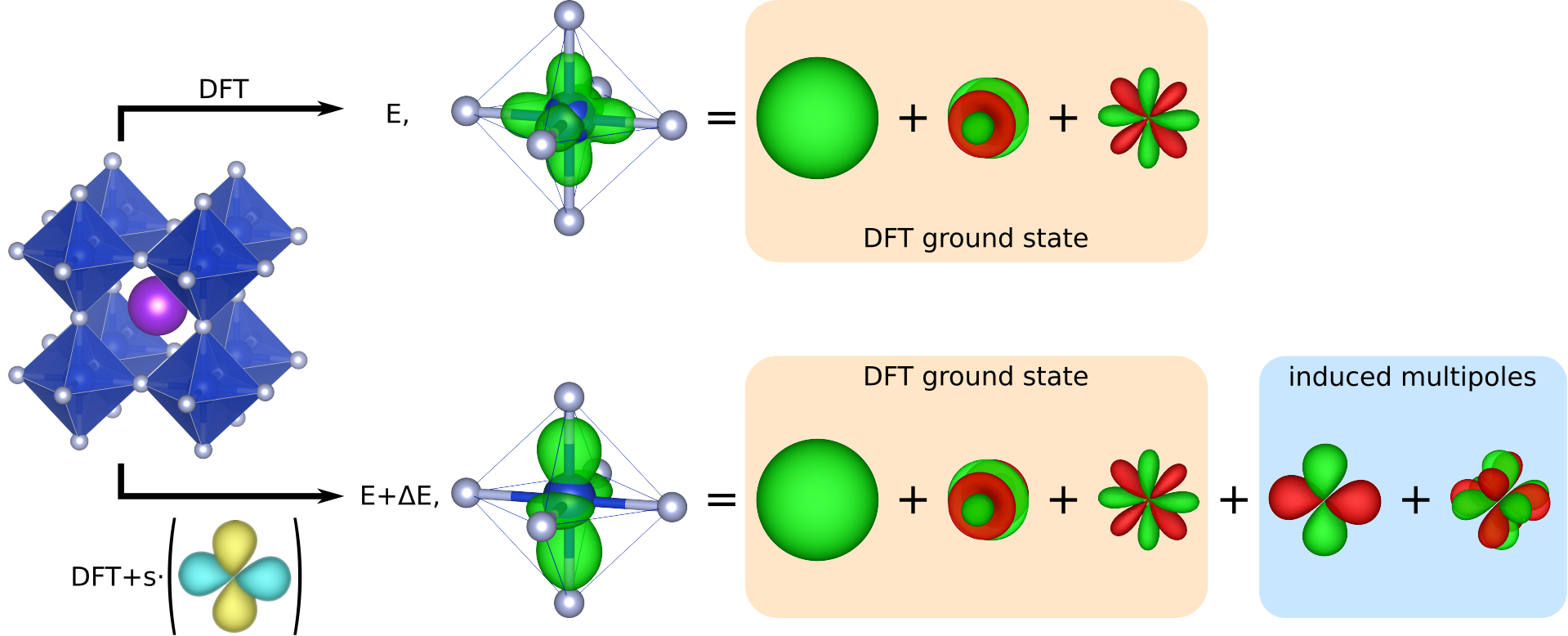}
    \caption{Visualization of our method for applying potential shifts to manipulate multipoles. (Top) Starting from a crystal structure, DFT normally computes the ground-state energy $E$ and the corresponding density, which can then be decomposed into multipoles.
    (Bottom) By applying an orbital-dependent potential shift $s$, the charge density is changed by directly induced multipoles of the same symmetry as the shift and possibly indirectly induced multipoles with different symmetries. The change in energy $\Delta E$ reveals the energetics of the obtained charge density.}
    \label{fig:schematic_method}
\end{figure*}

Now, we present the formalism we use to control or induce multipoles by applying a suitable local potential shift.
We start within the framework of constrained DFT, in which a constraint on the density is implemented by adding this constraint to the energy functional using the method of Lagrange multipliers \cite{dederichs_ground_1984}. To constrain some or all charge multipoles $w^I_{kt}$ for a given angular momentum $l$ to the desired values $\tilde w^I_{kt}$, one uses
\begin{align}
    E[\{\tilde w^I_{kt}\}] = \min_{n(\bm r), s^I_{kt}} ( & E_\mathrm{DFT} [n(\bm r)] \nonumber\\
    & - \sum_{Ikt} s^I_{kt} \left(w^I_{kt} [n(\bm r)] - \tilde w^I_{kt} \right) ) ,
    \label{eq:dederichs}
\end{align}
where $E_\mathrm{DFT}$ is the original DFT (or DFT+$U$) functional, $s^I_{kt}$ are the Lagrange multipliers, and the sum runs over the constrained multipoles.

Performing the minimization with respect to $n(\bm{r})$ leads to the usual Kohn-Sham equations but with an additional contribution to the Kohn-Sham potential proportional to $s^I_{kt} \mu^{kt}_{m'm}$ resulting from the constraint ({\it cf.} \pref{eq:shift_dft}).
In practice, finding a solution that satisfies the constraint for specific $\tilde{w}^I_{kt}$ then requires self-consistent adjustment of the Lagrange multipliers. However, since in most situations, it is of greater interest to systematically vary the multipole moments within a certain range rather than constraining them to a specific value, we use a Legendre transformation $\mathcal E := E - \sum_{Ikt} s^I_{kt} \tilde w^I_{kt}$ to change from $\tilde w^I_{kt}$ to $s^I_{kt}$ as the independent variables, analogous to the procedure in Refs.~\onlinecite{pickett_reformulation_1998, cococcioni_linear_2005}:
\begin{align}
    \mathcal E[\{s^I_{kt}\}] = \min_{n(\bm r)} ( E_\mathrm{DFT} [ n(\bm r) ]
    - \sum_{Ikt} s^I_{kt} w^I_{kt} [ n(\bm r) ] )
    \label{eq:cococcioni}
\end{align}

As already stated above, minimization of the first term on the right-hand side of \pref{eq:cococcioni}, i.e., $ E_\mathrm{DFT}$, leads to the usual Kohn-Sham equations. The second term results in an additional contribution to the Kohn-Sham equations, which can be obtained by taking the derivative with respect to $\langle \psi_{\bm p\nu}|$. Using \pref{eq:dens_to_mult} and \pref{eq:dens_in_dft}, this becomes
\begin{align}
    \frac\partial{\partial \langle \psi^*_{\bm p\nu}|} &\left( - \sum_{Ikt} s^I_{kt} w^I_{kt} \right)
    = -\sum_{Ikt} s^I_{kt} \sum_{mm'} \frac{\partial w^I_{kt}}{\partial \rho^I_{mm'}} \frac{\partial \rho^I_{mm'}}{\partial \langle \psi^*_{\bm p\nu}|} \nonumber\\
    &= \sum_{Imm'} \underbrace{\left( -\sum_{kt} s^I_{kt} \mu^{kt}_{m'm} \right)}_{:= \delta V^I_{mm'}} f_{\bm p\nu} P^I_{mm'}\ket{\psi_{\bm p\nu}} .
    \label{eq:shift_dft}
\end{align}
This defines the additional orbital-dependent contribution to the Kohn-Sham potential $\delta V^I_{mm'} P^I_{mm'}$. The additional potential thus acts only on the $l$ component of the wave functions at sites $I$ and consists of a sum of ``potential shifts'' of magnitude $s^I_{kt}$ that have an orbital $mm'$ dependence corresponding to the multipole operators $\mu^{kt}_{m'm}$.

The bottom panel of \pref{fig:schematic_method} illustrates the effect of such an additional potential contribution, in this case for an $(\xxyy)$-type charge quadrupole, which then per \pref{eq:cococcioni} induces multipoles, either directly with the same $(\xxyy)$ symmetry or indirectly with a different symmetry, such as the $(x^2-y^2)z^2$ hexadecapole shown.
In practice, we are usually constraining only one specific $l$ component of the corresponding multipole moment, as defined in \pref{eq:dens_to_mult} and \pref{eq:definition_mu}. 

To obtain the correct energy $E[\{\tilde w^I_{kt}\}]$ of the constrained system, which equals the DFT energy $E_\mathrm{DFT}[n(\bm r)]$ evaluated on the constrained charge density, one also has to consider the additional potential contribution to the Kohn-Sham equations.
Typically, in the calculation of total energy, the kinetic-energy contribution of the Kohn-Sham system is computed as the band energy, i.e., the sum over all occupied Kohn-Sham eigenvalues, minus the potential energy of the Kohn-Sham system in the effective potential (see, e.g., Ref.~\onlinecite{shick_implementation_1999}).
Therefore, with the additional contribution to the Kohn-Sham potential $\delta V^I_{mm'} P^I_{mm'}$, this potential energy increases by
\begin{align}
    \sum_{\bm p\nu} f_{\bm p\nu} &\expval{\sum_{Imm'} \delta V^I_{mm'} P^I_{mm'}}{\psi_{\bm p\nu}}
    = \sum_{Imm'} \delta V^I_{mm'} \rho^I_{mm'} .
    \label{eq:potential_energy}
\end{align}
Thus, the contribution from \pref{eq:potential_energy} has to be subtracted from the total energy in the DFT code.

Alternatively, the total energy can be obtained by taking $\partial E[\{\tilde{w}^I_{kt}\}]/\partial \tilde{w}^I_{kt}$ in \pref{eq:dederichs}, using the Hellmann-Feynman theorem, and then integrating over the applied potential shift (see also Ref.~\onlinecite{dederichs_ground_1984}):
\begin{align}
    \frac{\partial E[\{\tilde{w}^I_{kt}\}]}{\partial \tilde w^I_{kt}} &= s^I_{kt}
    \label{eq:hf_diff} \\
    \Rightarrow\quad
    E[\{\tilde w^I_{kt}\}] - E[\{ 0 \}] &= \int_{\{0\}}^{\{\tilde w^I_{kt}\}} \sum_{Ikt}  s^I_{kt}[\tilde w'^I_{kt}] \diff \tilde w'^I_{kt} \quad,
    \label{eq:hf}
\end{align}
where $\{0\}$ represents the multipoles of the reference state without the additional potential applied.
In \pref{sec:instability}, we use \pref{eq:hf_diff} to understand the hysteresis behavior of quadrupoles and their instabilities, whereas \pref{eq:hf} serves as a consistency check of the implementation of the energy calculation in our DFT modification.

\subsection{DFT details}

We implemented the method outlined in the previous subsection, including the additional potential terms and the necessary modifications to the total energy calculation in a modified version of the ``Vienna Ab-initio Simulation Package'' (VASP) in version 6.3.0 \cite{kresse_ab_1993, kresse_efficient_1996}.
The necessary python code is publicly available and the modifications to the VASP code are documented on GitHub \cite{maximilian_e_merkel_2023_8199391}.
Our modified VASP version is used for all DFT calculations presented in this work.

The calculations presented here are performed using a cubic $2\times2\times2$ supercell containing 8 formula units of \KCF or \KCZF.
We employ the PBE exchange-correlation functional \cite{perdew_generalized_1996} and projector augmented-wave (PAW)-type potentials~\cite{blochl_projector_1994, kresse_ultrasoft_1999} for K, Cu, Zn, and F as provided by VASP, where the semicore $3p$ states for K are included in the valence.
Except where otherwise noted, the strong electron-electron interaction in the Cu $3d$ orbitals are corrected with an on-site Hubbard interaction $U_\mathrm{eff}$ \cite{dudarev_electron-energy-loss_1998}, with a value of \SI{6.6}{eV}~\cite{liechtenstein_density-functional_1995}.
Lattice constants are obtained from relaxations in the ideal cubic perovskite structure without spin polarization or the $+U$ correction and are $a = \SI{4.092}{\angstrom}$ for \KCF and $a = \SI{4.136}{\angstrom}$ for \KZF. For \KCZF, we linearly interpolate the lattice constant between these values.
We use a $5\times5\times5$ and a $7\times7\times7$ $k$-point grid centered at the $\Gamma$-point for calculations with and without relaxations, respectively. The energy cutoff of the plane-wave basis set is chosen as \SI{900}{eV} to ensure convergence of higher-order multipoles.
The energy convergence criterion is set to \SI{e-8}{eV}, and the tetrahedron method is used to perform Brillouin-zone integrations.

We note that the +$U$ correction, the multipole calculation, and the multipole-shifting potential all use the same PAW projectors.
Furthermore, in this manuscript, we only apply potential shifts on quadrupoles of Cu-$3d$ orbitals and we generally switch off symmetries for electronic relaxations in the DFT calculations unless explicitly stated, to allow the charge density to break the crystal symmetry, either spontaneously or due to applied potential shifts.

\section{Results}

In this section, we present the results of our constrained multipole calculations for \KCF and its solid solutions with \KZF. 
We focus on charge quadrupoles, with $k=2$, and use the usual Cartesian representations of the multipoles shown in \pref{fig:crystal_struct_quads}(a), $Q_{z^2} \equiv w_{20}$, $Q_{x^2-y^2} \equiv w_{22}$.

\subsection{Exploring the electronic instability of \KCF}
\subsubsection{\KCF without electronic instability ($U=0$)}

As a first application of our method, we demonstrate that a potential shift indeed induces a multipole with the same symmetry as the applied shift and that we can control the magnitude of the multipole by varying the amplitude of the shift.
To prove these capabilities in a simple system, we study cubic \KCF using nonmagnetic DFT calculations with $U_\text{eff} = 0$ and apply an $(x^2-y^2)$-quadrupolar shift (i.e., $k=2$ and $t=2$) on the Cu sites. The absolute amplitude of this shift, $s_{x^2-y^2}$, is the same on every site, but the sign alternates to create a G-type antiferro-quadrupolar order.
We choose this type of order because its symmetry is the same as that of the experimentally observed Jahn-Teller distortion in \KCF.

\begin{figure}
    \centering
    \includegraphics[width=1\linewidth]{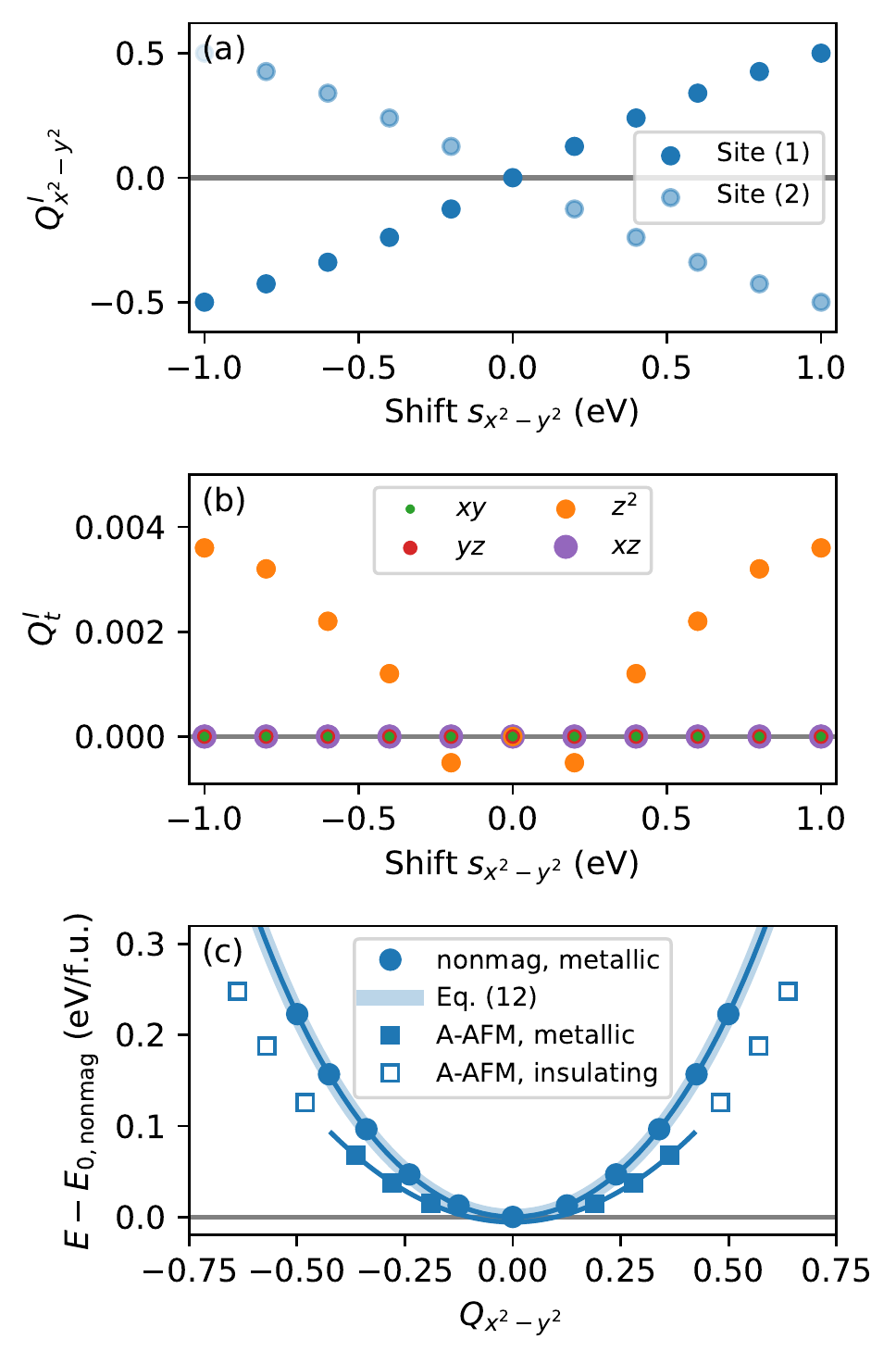}
    \caption{G-type antiferro-$Q_{x^2-y^2}$-type quadrupolar order in structurally cubic \KCF at $U=0$ obtained by applying a G-type antiferroic shift with amplitude $s_{x^2-y^2}$. (a) \xxyy and (b) other quadrupoles on both types of sites $I$ as a function of the applied shift $s_{x^2-y^2}$.
    (c) Energies as a function of $Q_{x^2-y^2}$ from nonmagnetic DFT (circles), \pref{eq:hf} based on the Hellmann-Feynman theorem applied on the data in (a) (thick line), and spin-polarized DFT with A-type antiferromagnetic order (squares). The DFT energies of the metallic regime (nonmagnetic and spin-polarized with $|Q_\xxyy| < 0.4$) are fitted with even polynomials (thin lines).
    }
    \label{fig:U0}
\end{figure}

\pref{fig:U0}(a) shows the quadrupoles $Q^I_{x^2-y^2}$ on the two sublattices obtained from the DFT calculations as functions of the shift amplitude $s_\xxyy$. For $s_{x^2-y^2}=0$, we recover the cubic \Pmtm ground-state charge density without any quadrupoles. For a finite potential shift, $|s_{x^2-y^2}|>0$, we induce $(\xxyy)$ quadrupoles with the same magnitude on each site but the sign of the quadrupole equals the sign of the applied shift, $Q_{x^2-y^2} := Q^{(1)}_{x^2-y^2} = -Q^{(2)}_{x^2-y^2}$. The induced quadrupoles are linear for small shifts and then start to saturate for larger shifts, with $Q^I_\xxyy[-s_\xxyy] = -Q^I_\xxyy[s_\xxyy]$, as mandated by symmetry.

The sizes of the other quadrupole components, corresponding to $t \neq 2$, are shown in \pref{fig:U0}(b). The $t_{2g}$-type quadrupoles $Q_{xy}$, $Q_{xz}$ and $Q_{yz}$ are numerically zero, while a small, ferro-ordered $Q_\zz$ develops due to its coupling to the $Q_\xxyy$ quadrupole. However, $Q_\zz$ always remains at least two orders of magnitude smaller than $Q_\xxyy$.
We note that in addition to the quadrupoles, some of the higher-order hexadecapoles ($k=4$) also respond to the $s_\xxyy$ shift. Therefore, this shift changes the magnitude of the hexadecapoles allowed in cubic symmetry and also induces an additional one due to the lowered symmetry.

The total energy as a function of the $Q_{x^2-y^2}$ quadrupole is shown in \pref{fig:U0}(c). First, we note that the energy obtained from the nonmagnetic DFT calculations has its minimum at $Q_{x^2-y^2}=0$. Therefore, the system does not exhibit an electronic instability toward a nonzero quadrupole moment. 
The energy curve yields a parabolic dependence for small $Q_{x^2-y^2}$ with a stiffness $\partial^2 E/\partial Q_\xxyy^2 = \SI{1.58}{eV}$.
Furthermore, the calculations always result in a metallic state. These results are consistent with earlier calculations showing that for $U=0$, \KCF exhibits no structural Jahn-Teller instability \cite{liechtenstein_density-functional_1995}.

As a consistency check, we also show in \pref{fig:U0}(c) the energy obtained from \pref{eq:hf}, i.e., from integrating a third-order fit of the $s[Q]$ curve shown in \pref{fig:U0}(a) with the zero-quadrupole ground state as reference energy. This integrated energy overlaps perfectly with the total energy calculated directly in the modified DFT code.

We then perform additional calculations in which we allow for spin polarization in the experimentally observed A-type antiferromagnetic order, i.e., with ferromagnetically ordered layers stacked antiferromagnetically along the [001] direction. The resulting total energy as a function of $Q_{x^2-y^2}$ is also shown in \pref{fig:U0}(c).
We were only able to stabilize this antiferromagnetic state for shifts $|s_\xxyy| \geq 0.2$ ($|Q_\xxyy| \gtrsim 0.19$). Thus, the presence of the quadrupolar order stabilizes the antiferromagnetic order (even without a $+U$ correction) relative to the nonmagnetic order. The resulting $Q^I_\xxyy$ have the same absolute value on each site, even though the relative sign between magnetic and quadrupolar moment varies from site to site due to their different spatial orderings, indicating a local biquadratic coupling in the lowest order.
Large quadrupoles, $|Q_\xxyy| > 0.4$, result in a transition in the electronic structure from metallic to insulating.

A parabolic fit of the energies in the metallic regime, i.e., for small but nonzero $Q_\xxyy$, results in a stiffness of \SI{1.11}{eV}. Thus, the stiffness is reduced compared to the nonmagnetic case, showing that the quadrupolar and magnetic order interact cooperatively. We also note that the parabolic fit to the antiferromagnetic total energy extrapolates to the same value at $Q_\xxyy = 0$ as the corresponding nonmagnetic energy (within the fitting accuracy), which suggests that for $Q_\xxyy=0$ the nonmagnetic and antiferromagnetic states are nearly degenerate.

These results illustrate that, even though for $U=0$ \KCF does not exhibit an electronic instability toward spontaneous quadrupolar order, our method allows us to systematically map out the corresponding energy surface, determine the stiffness with respect to quadrupole formation, and study, for example, how the formation of quadrupoles interacts with different types of magnetic order.

\subsubsection{\KCF with electronic instability ($U > 0$)}
\label{sec:instability}

Next, we investigate the spontaneous electronic instability of \KCF in a more realistic setting by including both magnetic order and the $+U$ correction in the calculations \cite{liechtenstein_density-functional_1995}. Again, we use our method of applying quadrupolar local potential shifts to systematically vary the local charge quadrupole in \KCF, independently of the structural degrees of freedom, i.e., with all atoms remaining in their positions corresponding to the ideal cubic perovskite structure.
As before, we apply an $(x^2-y^2)$-type shift with the same G-type antiferro-quadrupolar order as observed experimentally. In addition to the $Q_\xxyy$ quadrupole moment, we also monitor the $Q_\zz$ quadrupole. According to the nominal $d^9$ valence of the Cu$^{2+}$ cation, \KCF has one hole residing in the Cu-\eg bands. Depending on 
which specific linear combination of $e_g$ orbitals is occupied (or left empty) by this hole, this can create either a $Q_\xxyy$- or a $Q_\zz$-type quadrupole (or a linear combination thereof). The exact relation between hole orbital and quadrupole moment is discussed in \pref{app:quad_mix_angle}.
For the remainder of this manuscript, we focus on constraining $Q_\xxyy$, which dominates the orbital order in \KCF \cite{kugel_crystal_1973, pavarini_mechanism_2008}.

In all of the following calculations, we also employ a G-type antiferromagnetic order, which has an energy difference of less than \SI{8}{meV/}formula unit (\si{\fu}) compared to the experimentally observed A-type antiferromagnetic ground state. For our purposes, the G-type antiferromagnetic order has the advantage of conserving the local cubic site symmetry, and thus the orbital degeneracy of the Cu-$e_g$ states, while still introducing a local spin splitting that allows the system to become insulating.
The orbital degeneracy of the Cu-$e_g$ states allows us to enforce electronic cubic symmetry, which implies zero quadrupolar potential shifts, such that the system converges to a nonquadrupolar state, and thus to clearly separate charge effects and magnetic ordering.

\begin{figure}
    \centering
    \includegraphics[width=1\linewidth]{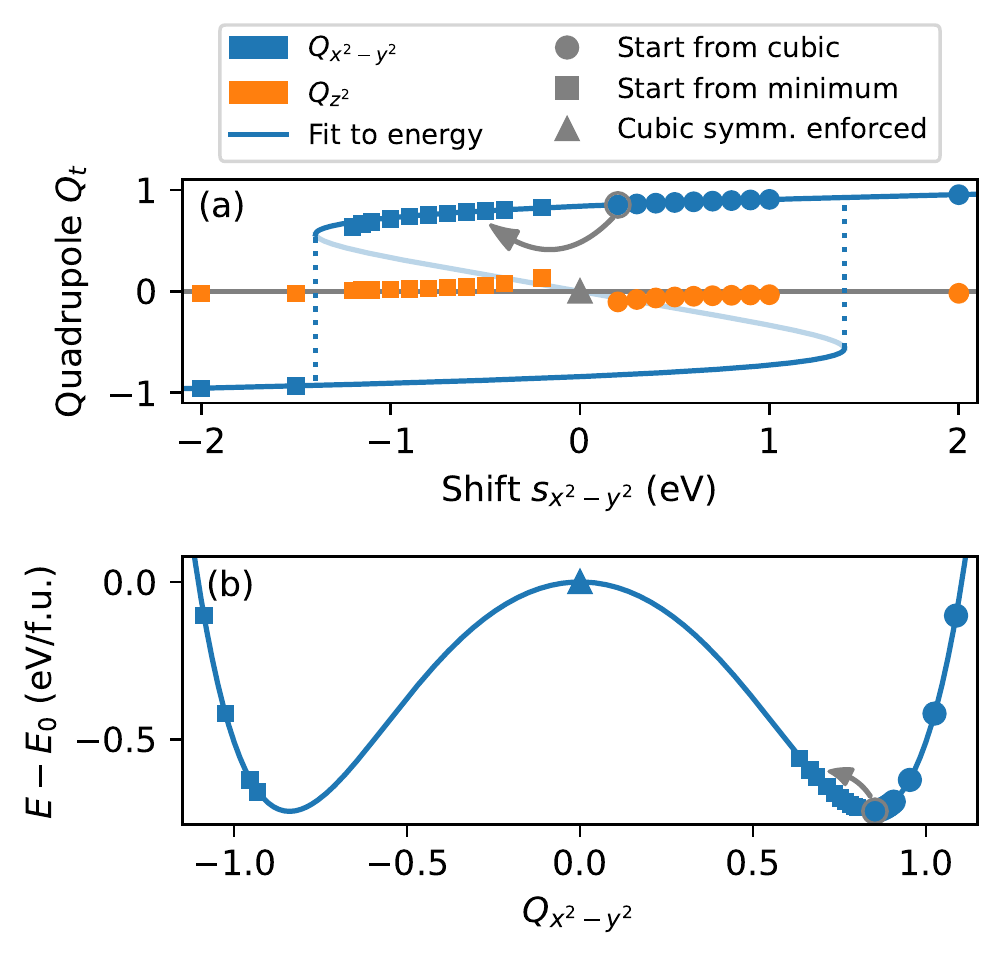}
    \caption{G-type antiferro  $Q_{x^2-y^2}$ quadrupolar order in cubic \KCF at $U=\SI{6.6}{eV}$ obtained by applying a shift $s_{x^2-y^2}$ with the same symmetry.
    Three types of calculations are shown: a zero-quadrupole calculation with cubic symmetry enforced (triangle), calculations with shifts starting from a charge density with cubic symmetry (circles), and calculations with negative shifts starting from the charge density for the smallest positive shift, as indicated by the arrows, to access the inner part of the energy well (squares).
    (a) $Q_{x^2-y^2}$ and indirectly induced $Q_{z^2}$ as a function of the applied shift $s_{x^2-y^2}$. 
    (b) Energy as a function of $Q_{x^2-y^2}$. The line is a sixth-order, even polynomial fit 
    $\SI{-1.53}{eV}\, Q_\xxyy^2 -\SI{0.06}{eV}\, Q_\xxyy^4 + \SI{1.08}{eV}\, Q_\xxyy^6$.
    The calculation with cubic symmetry enforced defines the reference energy $E_0$.
    }
\label{fig:U66_AF_order_energy}
\end{figure}

We first perform such a reference calculation enforcing cubic electronic symmetry. Consequently, the corresponding quadrupoles are zero (indicated by the gray triangle in \pref{fig:U66_AF_order_energy} (a)).
Then, we allow for an electronic symmetry breaking and apply a (positive) potential shift $s_{x^2-y^2}$. 
\pref{fig:U66_AF_order_energy} (a) shows the resulting $Q_\xxyy$ and $Q_\zz$ on one representative site as a function of $s_\xxyy$ (see data points marked as circles and labeled as ``start from cubic'').
Already for the smallest shift shown here, \KCF develops a strong quadrupole moment of $Q_\xxyy \approx 0.85$, which then increases only very gradually on further increasing $s_\xxyy$. This behavior is of course reminiscent of the spontaneous symmetry breaking, i.e., spontaneous quadrupole formation, already observed previously in structurally cubic \KCF \cite{liechtenstein_density-functional_1995}. 

Additionally, the symmetry breaking allows for a ferroically ordered $Q_\zz$ quadrupole, which, however, has a much smaller magnitude than $Q_\xxyy$ and becomes suppressed at larger shifts $s_\xxyy$.
For the remainder of this work, we therefore focus on the dominant quadrupole component $Q_\xxyy$.

The tendency for spontaneous quadrupole formation can also be seen from the total energy as function of $Q_\xxyy$, shown in \pref{fig:U66_AF_order_energy} (b). 
The symmetry breaking induced by a small positive shift, leads to an energy lowering of \SI{.73}{eV/\fu} compared to the energy $E_0$ of the nonquadrupolar high-symmetry calculation.
Increasing the potential shift then leads to a pronounced increase in energy. This allows to map out the ``outer'' part of the underlying energy surface, i.e., corresponding to values of the quadrupole moment that are larger than the equilibrium ground state value. Note that, within standard DFT($+U$) calculations, only the high-symmetry nonquadrupolar state and the low-symmetry ground state would be accessible. 

In order to also map out the ``inner'' part of the energy surface around its minimum, i.e., corresponding to (positive) values of $Q_\xxyy$ smaller than the ground state value, we perform calculations where we apply a negative potential shift, while starting from the converged charge density obtained with the smallest positive shift, i.e., close to the ground state, as indicated by the gray arrows in \pref{fig:U66_AF_order_energy}. 
The corresponding results are shown in \pref{fig:U66_AF_order_energy} as square symbols and labeled as ``start from minimum''. It can be seen that this initialization indeed allows us to converge the charge density into a local minimum of \pref{eq:cococcioni}, which results in a weak decrease of the $Q_\xxyy$ quadrupole (see \pref{fig:U66_AF_order_energy}(a)) and the corresponding increase in the total energy (see \pref{fig:U66_AF_order_energy}(b)). At a certain amplitude of the negative shift potential, $Q_\xxyy$ then jumps to a larger negative value, slightly lower than the negative equilibrium value.

Without an applied shift, the two oppositely polarized states, with positive or negative quadrupole moment on the selected site, are of course energetically degenerate. By applying quadrupolar potential shifts with the respective opposite sign, the system can be switched from one of the corresponding domain states to the other.
We can fit the calculated energies with an even sixth-order polynomial. The fit, indicated by the solid lines in \pref{fig:U66_AF_order_energy}, shows almost perfect agreement with the calculated values, with a maximum deviation of \SI{7}{meV/\fu}
It clearly shows that the energy as a function of $Q_\xxyy$ forms a symmetric ``double well'' with two degenerate minima around $Q_\xxyy =  \pm 0.84$.
The fit also allows us to estimate the stiffness of the energy around the minimum to be $\partial^2 E/\partial Q_\xxyy^2 = \SI{12.55}{eV}$. This stiffness is an important quantity to characterize fluctuations of $Q_\xxyy$ around its equilibrium value.

From this fit, we can also obtain a relation for the quadrupole moment as a function of the applied shift, $Q[s]$, by taking the derivative of $E[Q]$ and using \pref{eq:hf_diff} to obtain $s[Q]$. The resulting curve is also shown (as a solid blue line) in \pref{fig:U66_AF_order_energy} (a).
It correctly predicts hysteresis as a function of the shift due to the double-well potential and agrees almost perfectly with the explicitly calculated values, both in terms of the overall size of $Q[s]$ as well as with respect to the width of the hysteretic regime.
The quadrupoles obtained using positive shifts all lie on the upper branch of the hysteresis curve, whereas the values obtained for negative shifts first also follow the upper branch and then jump onto the lower branch for large negative shifts beyond the extremum of $s[Q]$ (or equivalently, the inflection point of $E[Q]$).

This example shows that our method allows us to systematically map out the energy landscape as a function of the quadrupolar order parameter around its equilibrium value for a system that exhibits spontaneous instability, within the given hysteretic limits.

\subsection{Intersite quadrupole-quadrupole interactions}
\label{sec:orbital_order}

\begin{figure}
    \centering
    \includegraphics[width=1\linewidth]{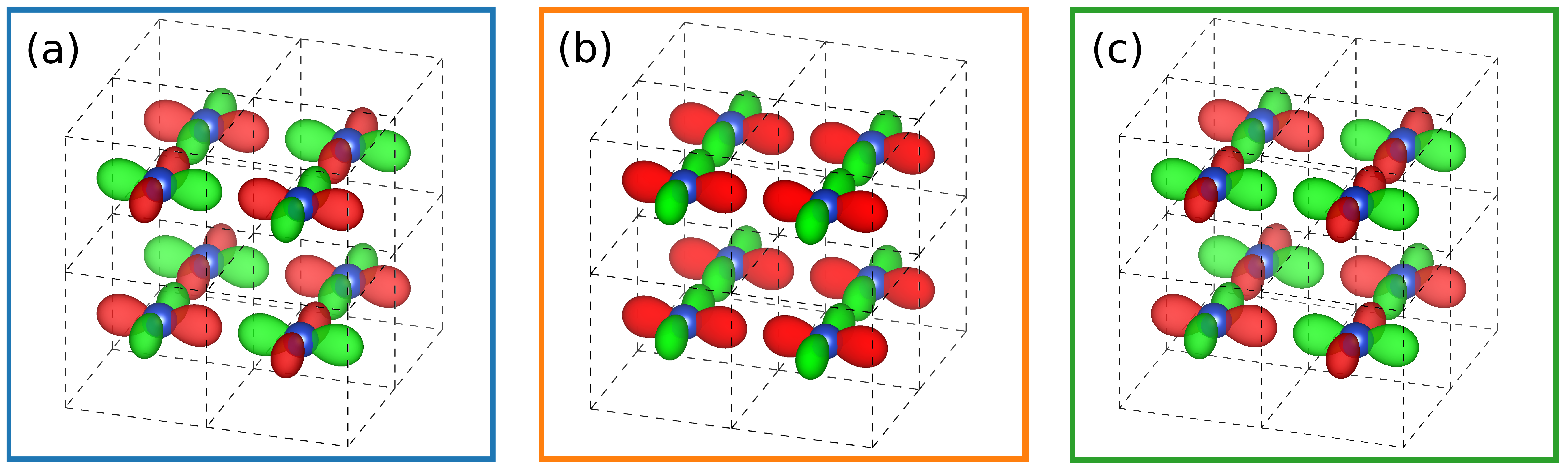}
    \includegraphics[width=1\linewidth]{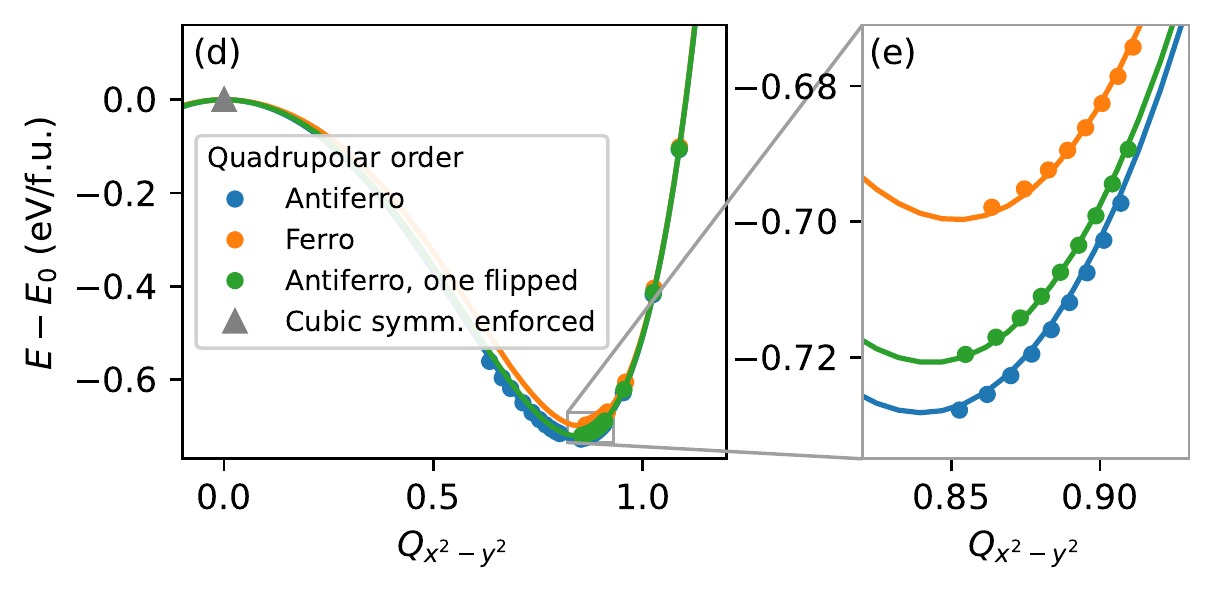}
    \caption{ (a,b,c) Schematic representation of different quadrupolar orderings in cubic \KCF. We consider (a) G-type antiferro-quadrupolar ordering, (b) ferro-quadrupolar ordering, and (c) G-type antiferro-quadrupolar ordering with one quadrupole flipped in sign. (d, e) Energy with respect to the zero-quadrupole energy $E_0$ as a function of the average absolute $|Q_{x^2-y^2}|$ for the quadrupolar orderings visualized in (a-c). The solid lines are even, sixth-order polynomial fits.
    }
\label{fig:KCuF3-ordering-fig}
\end{figure}

Up to now, all calculations were performed for the G-type antiferro-quadrupolar order, compatible with the experimentally observed antiphase Jahn-Teller distortion. Now, we explore the effect of different quadrupolar orders to calculate the strength and nature of intersite quadrupole-quadrupole interactions. We keep a G-type antiferromagnetic order for all these calculations.
\pref{fig:KCuF3-ordering-fig} (a-c) shows the different orders we consider: (a) the previously used G-type antiferro-quadrupolar order, (b) a ferro-quadrupolar order, and (c) the same as in (a) but with the sign of one multipole out of eight flipped.

We can stabilize these orders by applying small quadrupolar potential shifts with the appropriate site-dependent signs. Note that such states would be difficult to access without our new method.

We plot the energy as a function of the site-averaged absolute quadrupole $Q_\xxyy^I$ in \pref{fig:KCuF3-ordering-fig}(d, e). We note that configuration (c) breaks the symmetry between sites, but this symmetry breaking leads to only small differences of less than 0.02 between the $|Q_\xxyy^I|$ on different sites.
As can be seen in \pref{fig:KCuF3-ordering-fig}(d, e), all quadrupolar orders lead to an energy lowering of around \SI{.7}{eV/\fu}, relative to the nonquadrupolar reference state, whereas the energy differences between the different configurations are significantly smaller, around \SI{30}{meV/\fu} between the ferro- and antiferro-quadrupolar configuration.

Thus, we can distinguish two rather different energy scales. The energy lowering of \SI{.7}{eV/\fu} can be related to the local quadrupole formation, since this energy contribution is independent of the specific spatial ordering
\footnote{Strictly speaking, even-power higher-order, e.g., biquadratic, intersite couplings could also contribute to this energy lowering. However, from the results presented in \pref{sec:substitution} it can be seen that the energy lowering for an isolated Cu site in \KCZF is also around \SI{.7}{eV/Cu}, which implies that such higher order intersite couplings are negligible.}.
The second energy scale of around \SI{30}{meV/\fu}, defined by the energy differences between the different spatial arrangements of quadrupoles, is the energy scale of the intersite quadrupolar-quadrupolar interactions. The corresponding energy differences can in principle be mapped on an Ising-like model (potentially taking into account also energies of further spatial configurations to incorporate anisotropic or further-neighbor couplings of the quadrupoles). Such a model can then be used to estimate the hypothetical, purely electronic quadrupolar ordering temperature, or can be further extended by explicitly taking into account also structural degrees of freedom, e.g., a corresponding Jahn-Teller distortion.
Thus, our approach allows clear separation of the purely electronic interactions from electron-phonon-related or other structural contributions.

\subsection{Atomic substitution from \KCF to \KZF}
\label{sec:substitution}

Finally, we address the suppression of the quadrupolar tetragonal phase of \KCF by substitution of Cu with Zn. \KZF is a cubic perovskite with a full $3d^{10}$ shell and experimentally, \KCZF becomes cubic at a concentration of $x > 0.4$, with a lattice constant of around \SI{4.06}{\angstrom} \cite{tatami_orbital_2007}. 
However, Monte-Carlo and cluster-expansion studies showed that even in the cubic phase, the remaining \CuFoct octahedra still exhibit a Jahn-Teller distortion, but without any long-range order \cite{tatami_orbital_2007, tanaka_randomly_2005}.
In the following, we use our method of constraining local quadrupoles to study the role of the electronic instability in cubic \KCZF, and to further characterize the nature of the tetragonal-to-cubic phase transition upon ionic substitution.
This is intended as a demonstration of the capabilities of our method rather than an exhaustive analysis of \KCZF.

We consider $x \in \{0, 0.5, 0.75, 0.875\}$, i.e., 8, 4, 2, and 1 Cu atom in the cubic $2\times2\times2$ supercell, respectively. For each case, we use the most symmetric arrangement of Cu and Zn, which allows us to converge a nonquadrupolar state by enforcing the respective crystal symmetry in our calculations. 
In these arrangements, the \CuFoct octahedra are always as far apart as possible and in particular there are no \CuFoct octahedra neighbors for $x \geq 0.5$.
We explicitly checked for $x=0.75$ that the ground-state energy difference to all other non-symmetry-equivalent configurations is only on the order of a few \si{meV/\fu} (here and in the following, f.u. always refers to one five-atom unit).

We keep the directions of the magnetic dipole moments on the Cu sites the same as in the previously studied G-type antiferro-magnetic state. 
However, since the Zn$^{2+}$ cations have a completely filled $3d$ shell, they do not exhibit any magnetic moment, which means that for $x \geq 0.5$, the chosen arrangement of Cu$^{2+}$ ions corresponds to ferromagnetic order on the Cu sublattice. The same applies to the G-type antiferro-ordered potential shifts.

\begin{figure}
    \centering
    \includegraphics[width=1\linewidth]{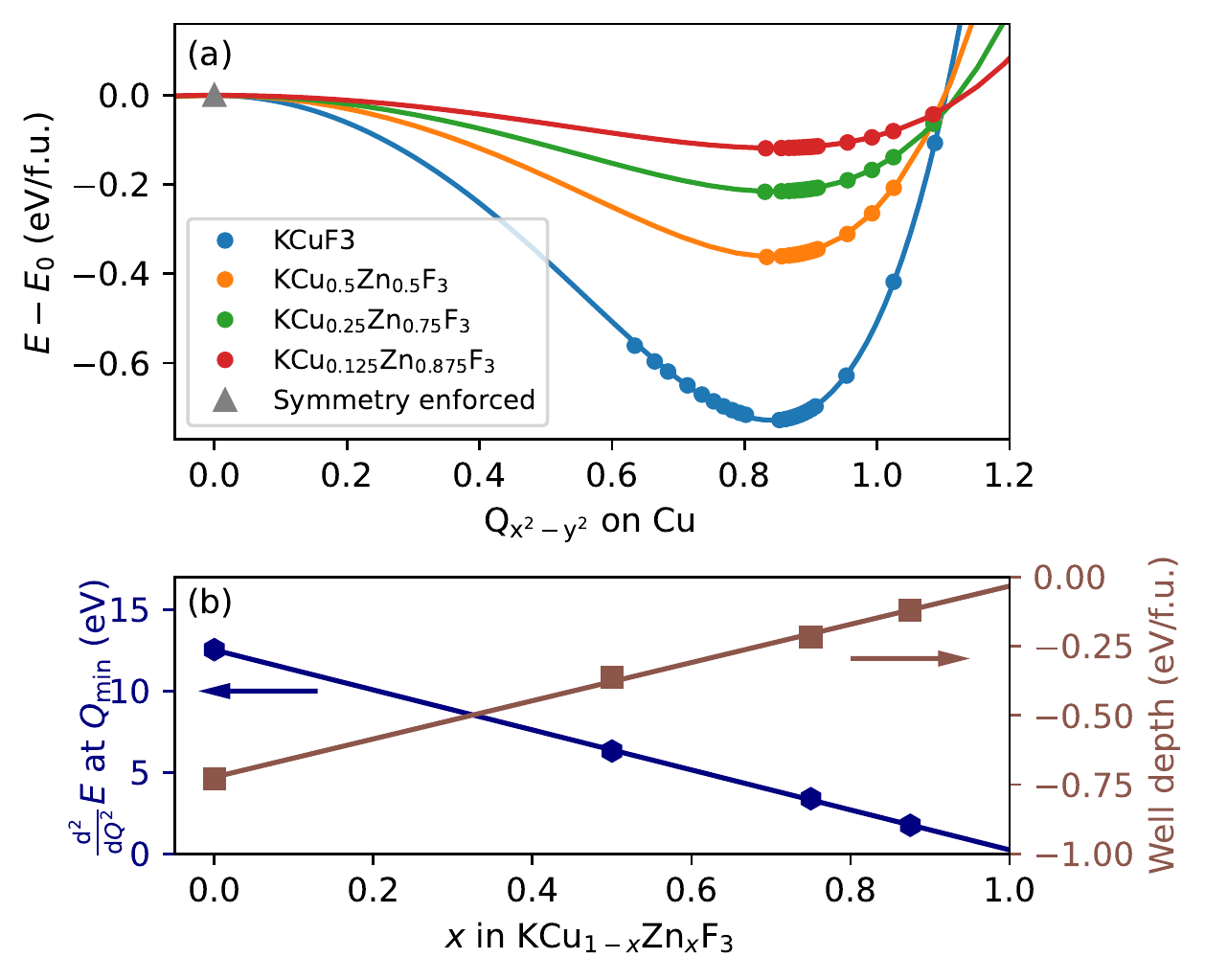}
    \caption{
    (a) Energy of \KCZF when inducing quadrupoles on the Cu sites for different Zn concentrations $x$.
    (b) Well depth and stiffness as a function of $x$.
    }
    \label{fig:substitution}
\end{figure}

As \pref{fig:substitution} (a) shows, quadrupole formation always leads to an energy lowering for all compositions we consider here, with a smaller energy lowering for higher Zn concentration $x$.
This persistence of the electronic instability shows that the $Q_\xxyy$ quadrupole will always tend to form locally on Cu sites, independent of the neighboring octahedra, which is consistent with the presence of locally distorted \CuFoct octahedra even in the experimentally reported globally cubic structure with $x>0.4$ \cite{tatami_orbital_2007, tanaka_randomly_2005}.

Again, we can reliably fit the data for all $x$ with a sixth-order, even polynomial. This allows us to extract the curvature and depth of the energy minimum, as depicted in \pref{fig:substitution} (b).
Both quantities scale almost perfectly linearly with the number of Cu atoms and extrapolate to zero for pure \KZF. This shows that the energy gain per Cu is basically independent of the concentration $x$.
These results strengthen the observation from  \pref{sec:orbital_order} that quadrupole formation is primarily a local on-site property.

\section{Summary and Conclusions}

In this work, we have presented a method to perform DFT calculations with constrained local charge multipole moments.
We implement this in the framework of constrained DFT using Lagrange multipliers \cite{dederichs_ground_1984}, which introduces orbital-dependent potential shifts coupling directly to specific multipoles. This allows us to vary both magnitude and spatial order of the local multipole moments and thereby disentangle electronic from structural degrees of freedom.
By enabling the exploration of the multipolar energy landscape also outside of local minima, our method represents a significant advance compared to previous methods based on, e.g., initialization of the density matrix.
We note that specific quadrupolar moments can straight-forwardly be constrained by iteratively adjusting the potential shifts with a script wrapping around our implementation. Alternatively, the Lagrange parameter required to obtain a specific quadrupole can directly be extracted from the systematic variation of shifts, as presented above in Figs.~\ref{fig:U0}(a) and \ref{fig:U66_AF_order_energy}(a).

While we envision many different applications, we have demonstrated the effectiveness of the method in disentangling the electronic instabilities of Jahn-Teller-active materials from their crystal structures by mapping out the charge-quadrupolar energy surface of \KCF. Our method enables access to the energetics of the material as a function of the charge quadrupole moments in the absence ($U = 0$) and presence ($U>0$) of electronic instabilities. For $U = 0$, we observed that the charge quadrupole order stabilizes an antiferromagnetic order. For the case of $U > 0$, the electronic instability leads to two degenerate energy minima and hysteretic behavior of the quadrupole moments as a function of the applied potential shift, which can be explored by using appropriate initialization. Such access to the energy as a function of the multipole moment for a fixed crystal structure is invaluable for understanding purely electronic as well as coupled electronic and structural phase transitions.

Furthermore, our method allows us to extract multipole-multipole interactions, which can be used to construct simplified Ising-like Hamiltonians for further studies by stabilizing various configurations with different signs of the local multipole moments on different sites.
In our \KCF example, we found that the scale of energy differences between the different quadrupolar configurations is around \SI{30}{meV/\fu}, similar to the purely electronic Kugel-Khomskii temperature obtained from earlier DFT+DMFT calculations \cite{pavarini_mechanism_2008}.
Therefore, by considering only the electronic degrees of freedom, our results further support the conclusion that structural distortions are vital for stabilizing the orbitally ordered state in \KCF to higher temperatures.

Finally, we showed that our method can be used to study the evolution of electronic instabilities in solid solutions by applying it to the example of \KCZF. Our analysis showed that the instability towards the formation of a local quadrupole moment on the Cu site persists down to low Cu concentrations, even where the average crystal structure is experimentally known to be cubic. The energy gain from the local quadrupole formation is around \SI{0.7}{eV/Cu} and thus much larger than the energy scale of the quadrupolar inter-site interactions (around \SI{30}{meV/\fu}).

While here we have demonstrated the capabilities of our multipole-constrained DFT method for the case of charge quadrupoles, the extension to other (higher-order) charge and magnetic multipoles is rather straightforward.
We anticipate, therefore, that the method will be invaluable in disentangling the contributions of coupled and/or competing magnetic, electronic, and lattice orders in a broad range of correlated materials \cite{dagotto_complexity_2005}, such as the pseudo-gap phase of unconventional superconductors~\cite{varma_mind_2010}, 5$d^1$~\cite{mansouri_tehrani_untangling_2021} and 5$d^2$~\cite{fiore_mosca_modeling_2022} transition-metal double perovskites, $f$-electron systems with hidden high-order multipolar order parameters~\cite{pourovskii_hidden_2021} and magnetoelectric multiferroics~\cite{van_aken_observation_2007, ederer_towards_2007}.

\appendix*
\begin{acknowledgments}
This work was funded by the European Research Council under the European Union’s Horizon 2020 research and innovation program project HERO (Grant No. 810451) and by ETH Z\"urich.
This work was supported by a grant from the Swiss National Supercomputing Centre (CSCS) under project ID s1128.
\end{acknowledgments}

\appendix

\section{Formulas for the multipolar operator}
\label{app:multipoles_definition}

In order to convert from the complex spherical harmonics used in Ref.~\onlinecite{bultmark_multipole_2009} to real spherical harmonics that are more intuitive to interpret, we use the following transformation from the shift matrix in spherical harmonics $\hat \mu$ to the corresponding matrix in real harmonics $\mu$:
\begin{align}
    \mu^{kt}_{mm'} &= \begin{cases}
        (\hat\mu^{k,t}_{mm'} - (-1)^t \hat\mu^{k,-t}_{mm'}) \mathrm i / \sqrt2 & t<0 \\
        \hat\mu^{k,0}_{mm'} & t=0 \\
        (\hat\mu^{k,-t}_{mm'} + (-1)^t \hat\mu^{k,t}_{mm'}) / \sqrt2 & t>0
    \end{cases}
    \label{eq:sph_to_real}
\end{align}
The multipoles $w^I_{kt}$ from \pref{eq:dens_to_mult} are real because the $\mu^{kt}$ are Hermitian matrices.

Since we only consider charge multipoles here, we are able to use a simpler notation than that of Ref.~\onlinecite{bultmark_multipole_2009}, with the complex-valued $\hat w_{kt}$ from \pref{eq:dens_to_mult} of this manuscript corresponding to the $w^{k0k}_t$ from Eq.~(26) in Ref.~\onlinecite{bultmark_multipole_2009}.

The transformation from density matrix to multipoles is a linear, invertible operation because the $\mu^{kt}$ form an orthogonal, complete basis for Hermitian $(2l+1)\times(2l+1)$ matrices. The completeness relations for the matrices $\mu$ are
\begin{align}
    \sum_{mm'} \mu^{kt}_{mm'} \mu^{k't'}_{m'm} &= \delta_{kk'} \delta_{tt'} / c^k \nonumber\\
    \sum_{kt} c^k \mu_{mm'}^{kt} \mu_{n'n}^{kt} &= \delta_{mn} \delta_{m'n'},
    \label{eq:completeness_multipoles}
\end{align}
with $c^k = (2k+1) n_{lk}^2$. Similar completeness relations also exist for $\hat\mu$.

\section{Connecting quadrupoles and the \eg-orbital mixing angle}
\label{app:quad_mix_angle}

Here, we present the relation between local quadrupoles and the \eg-orbital mixing angle in \KCF, which we exemplify by the data from \pref{sec:instability}.

\begin{figure}
    \centering
    \includegraphics[width=1\linewidth]{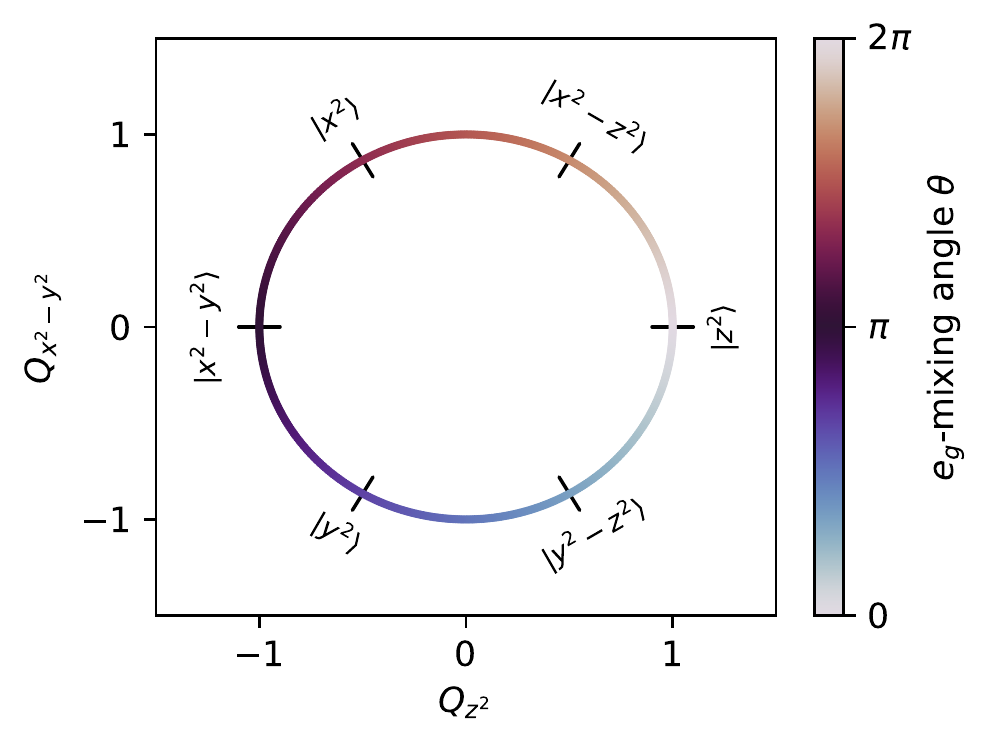}
    \caption{Quadrupoles $Q_\zz$ and $Q_\xxyy$ calculated for a $d^9$ shell where the hole occupies a linear combination of the \eg orbitals, parametrized by a mixing angle $\theta$. The possible quadrupole values lie on the unit circle, with the corresponding phase angle determined by $\theta$. }
    \label{fig:quad_kk}
\end{figure}

Insulating \KCF nominally has one hole occupying a Cu-\eg state. This state can be expressed as a linear combination of the $d_\zz$ and $d_\xxyy$ orbitals through an orbital-mixing angle $\theta$ \cite{khomskii_transition_2014}:
\begin{align}
\label{eq:mixing-angle}
    \ket\theta = \cos\frac\theta2 \ket\zz + \sin\frac\theta2 \ket\xxyy
\end{align}

If we analytically calculate the density matrix $\rho_{mm'} = \braket{m}{\theta} \braket{\theta}{m'}$ and the quadrupole operators $\mu^{2t}_{mm'}$, we obtain the quadrupoles through \pref{eq:dens_to_mult}:
\begin{align}
    Q_\zz &= \cos\theta \nonumber\\
    Q_\xxyy &= -\sin\theta \nonumber\\
    Q_{xz} &= Q_{yz} = Q_{xy} = 0
\end{align}
Therefore, the quadrupole moments lie on a circle with radius 1 in the $Q_\zz$-$Q_\xxyy$ plane, where the phase angle can be directly related to the \eg-mixing angle of the corresponding hole orbital, \pref{eq:mixing-angle}.
This relationship is visualized in \pref{fig:quad_kk}.

In a real material, the quadrupoles can deviate from that picture. For example, hybridization with ligands influences the occupation of \eg and \ttg orbitals, which can distort the circle or even create other quadrupoles. The radius, for example, corresponds to the \eg hole occupation and is therefore directly affected by hybridization.
Also, if the density matrix of the hole does not represent a pure state but a mixed state, the quadrupoles cannot be described by a single parameter $\theta$ anymore. This can lead to quadrupole values in the inner part of the circle.

In this context, we can now discuss the results shown in \pref{fig:U66_AF_order_energy}(a). For the smallest \xxyy potential shift, we obtain a large, antiferro-ordered $Q_\xxyy = \pm 0.85$ and a small, ferroic $Q_\zz = - 0.10$.
This indicates a hole occupation of $(Q_\xxyy^2 + Q_\zz^2)^{1/2} = 0.86$ and a mixing angle $\theta = \mp \ang{96.9}$.
The deviation from \ang{90}, which is the angle favored by the \xxyy shift, shows the influence of effects intrinsic to the material, such as superexchange. This influence gets further suppressed for larger shifts, where $\theta \rightarrow \mp\ang{90}$.
In principle, by applying different linear combinations of \xxyy- and \zz-type potential shifts, our method would also allow to map out the energy landscape as function of the orbital mixing angle. This could then give insights into the underlying superexchange mechanism.

\bibliography{bibfile}

\end{document}